\documentclass[12pt,a4paper]{article}            
 \usepackage[skins,theorems]{tcolorbox}
\tcbset{highlight math style={enhanced,
  colframe=red,colback=white,arc=0pt,boxrule=1pt}}
  \usepackage[bookmarksopen, bookmarksnumbered, bookmarksopenlevel=2]{hyperref}
  \usepackage{tikz}
  \usepackage{tikz-3dplot}
 \usetikzlibrary{calc}

 \usetikzlibrary{decorations}
 \usepackage[UKenglish]{babel}
 \usepackage[toc,page]{appendix}
 \usepackage{amsmath}
 \usepackage{amssymb}
 \usepackage{float}
 \usepackage{amsthm}
 \usepackage{graphicx}
 \usepackage{hhline}
 \usepackage[bf]{caption}
\usepackage{cite}
\usepackage[vcentermath]{youngtab}
\usepackage{geometry}
\usepackage{slashed}
\usepackage{color}
\usepackage{stackrel}
\usepackage{tikz-cd} 
\usepackage{mathtools}
\usepackage{cancel} 
\usepackage{multirow}
\usepackage[margin=0pt,font=small,labelfont=normalfont,skip=22pt]{subcaption}
\usepackage{empheq}
\usepackage{arydshln}
\usepackage{fancyvrb}
\usepackage{color,soul}

 \makeatletter
\g@addto@macro\bfseries{\boldmath}
\makeatother

\newcommand{\bqa}{\begin{eqnarray}}
\newcommand{\eqa}{\end{eqnarray}}

\newenvironment{eqn*}{\begin{equation*}\begin{aligned}}{\end{aligned}\end{equation*}\noindent}
\hypersetup{
    pdftitle={},
    pdfauthor={},
    pdfsubject={}
}

\numberwithin{equation}{section}
\numberwithin{table}{section}

\makeatletter

\newcommand{\be}{\begin{equation}}
\newcommand{\ee}{\end{equation}}
\newcommand{\beq}{\begin{equation}}
\newcommand{\eeq}{\end{equation}}
\newcommand{\ba}{\begin{aligned}}
\newcommand{\ea}{\end{aligned}}

\newcommand{\bea}{\begin{eqnarray}}
\newcommand{\eea}{\end{eqnarray}}

\newcommand\bi{\begin{itemize}}
\newcommand\ei{\end{itemize}}

\def\unit{{1\kern-.65ex {\rm l}}}
\def\1{{1\kern-.65ex {\rm l}}}

\def\dd{{\mathrm{d}}}

\newcount\hour \newcount\minute
\hour=\time \divide \hour by 60
\minute=\time
\def\now{%
\ifnum \hour<13
  \ifnum \hour=0 \advance \hour by 12 \number\hour:\else \number\hour:\fi%
     \ifnum \minute<10 0\fi%
     \number\minute%
\ A.M.%
\else \advance \hour by -12 \number\hour:%
  \ifnum \minute<10 0\fi%
  \number\minute%
  \ P.M.%
\fi%
}

\makeatother

\begin{document}

\begin{titlepage}
\begin{center}
\rightline{\small }

\vskip 15 mm

{\large \bf
A species scale-driven breakdown of effective field theory in time-dependent string backgrounds } 
\vskip 11 mm

Alek Bedroya$^{1}$, Hayden Lee$^{2}$, and Paul Steinhardt$^{3}$
\vskip 11 mm
\small ${}^{1}$ 
{\it Princeton Gravity Initiative, Princeton University, Princeton, NJ 08544, USA} \\[3 mm]
\small ${}^{2}$
{\it Department of Physics and Astronomy, University of Pennsylvania, Philadelphia, PA 19104, USA}\\[3 mm]
\small ${}^{3}$
{\it Department of Physics, Princeton University, Princeton, New Jersey 08544, USA}

\end{center}
\vskip 17mm

\begin{abstract}
 We present a novel way in which effective field theory (EFT) can break down in cosmological string backgrounds depending on the behavior of the quantum gravity cutoff in infinite distance limits, known as the species scale $\Lambda_s$. Namely, EFT can break down if the species scale $\Lambda_s$ falls off so rapidly as the Friedmann-Robertson-Walker (FRW) scale factor grows from some initial value $a_i$ to some final value $a_f$ that the physical momentum of an initial Hubble-sized perturbation ${\cal O}(H_i^{-1})$ grows to exceed the species scale. For EFT to remain valid, a new condition $H_i \frac{a_i}{a_f} \ll \Lambda_{s,f}$ must hold, which is distinct from Trans-Planckian conditions discussed in the literature. Using the universal relation $\frac{\nabla m}{m} \cdot \frac{\nabla \Lambda_s}{\Lambda_s} = \frac{1}{d-2}$ in the infinite distance limits of moduli space where $m$ is the mass scale of the lightest tower and $\nabla$ measures variations with respect to the canonical metric on moduli space, we show that spatially flat FRW solutions in the string landscape violate this condition or at best marginally satisfy it. However, we find that sufficiently large negative spatial curvature always avoids a breakdown. To avoid EFT breakdown, we derive an upper bound on the duration of quasi-de Sitter expansion that classically evolves to decelerated expansion. Our bound is proportional to the Trans-Planckian Censorship Conjecture (TCC) bound, with the advantage that it applies to any FRW solution in the string landscape. Finally, we distinguish EFT breakdown from TCC violation, the latter being a quantum gravity constraint rather than an EFT limitation. {\it Perhaps our most surprising finding is that in any flat FRW solution that develops a weakly coupled string at future infinity the EFT inevitably breaks down.}


\end{abstract}

\vfill
\end{titlepage}

\newpage

\tableofcontents

\setcounter{page}{1}

\section{Introduction}

In effective field theory (EFT), we can envision configurations in which the EFT breaks down in a small region of space, described as a defect. However, we can still study such defects using EFT, as they couple to and interact with light states that can be collectively described by fields. Therefore, we can scatter light states from the asymptotic region of spacetime onto such objects. In fact, this way of characterizing physics is fundamental to quantum gravity and forms the basis for holography. One field is particularly crucial for ensuring a valid spacetime description at such asymptotics: the graviton. The field-theoretic description of the graviton is essential for defining spacetime and its asymptotics, which, in turn, are used to define holographic observables. Therefore, it is necessary that the EFT description of the gravitational action remain valid near some asymptotic region of spacetime to enable the formulation of holography in such settings.

Typically, an EFT is considered to break down when the wavelengths associated with the spatial variation of the fields become shorter and shorter until they surpass the field theory cutoff, leading to a breakdown of the EFT. However, in this work, we explore a different mechanism for the breakdown of EFT. This occurs when, as fields propagate over time, the cutoff itself evolves and decreases to the point where perturbations previously considered to be in the infrared (IR) regime surpass the cutoff scale. This represents a novel mechanism of EFT breakdown, warranting particular attention in quantum gravity. As we review in Section \ref{sec:2}, this phenomenon is ubiquitous in string theory, where scalar-field cosmologies often feature a time-dependent and decreasing quantum gravity cutoff $\Lambda_s$, also known as the species scale. The species scale controls the higher-order curvature corrections to the effective action \cite{vandeHeisteeg:2022btw}. Therefore, if the length scale of perturbations falls below $\Lambda_s^{-1}$, the field theory description of the metric---and consequently, the very notion of spacetime---breaks down in the asymptotic region of spacetime, leaving us with a spacetime that lacks an asymptotic time direction and no clear framework for formulating holography.

In Section \ref{sec:3}, we demonstrate that this condition leads to the inequality
\begin{align}
H_i \frac{a_i}{a_f} \ll \Lambda_{s,f} \,,
\end{align}
where $a_i$ and $a_f$ are the initial and final scale factors in an expanding FRW cosmology, $H_i$ is the initial Hubble parameter, and $\Lambda_{s,f}$ is the final species scale. We explain how this differs fundamentally from the so-called trans-Planckian issues \cite{Martin:2000xs,Brandenberger:2012aj} in cosmology, where an IR mode can be traced back in time to have trans-Planckian physical momentum. In fact, in Section \ref{sec:5}, we argue that trans-Planckian issues do not constitute genuine breakdowns of EFT. In doing so, we also correct the widespread misconception that the absence of breakdown is due to the emergence of modes in the Bunch-Davies vacuum. 

Guided by the examples of black hole and de Sitter backgrounds, where a well-behaved set of modes exists that do not exhibit any trans-Planckian spacetime variations, we show that in any scenario of accelerated expansion, a well-behaved set of modes exists that can be used to show that the EFT does not necessarily break down. In short, the \textit{good} modes have the property that when they are traced back in time, the reverse of the Hubble friction leads to an abundance of IR modes that smooth out the UV behavior. This effect cannot resolve the type of EFT breakdown we described earlier, which occurs by tracing the modes in the future direction where the modes are redshifted rather than the past direction where they are blueshifted.

We then apply the inequality $H_i \frac{a_i}{a_f} \ll \Lambda_{s,f}$ to FRW spacetimes that arise in the weakly coupled corners of the string theory landscape. Perhaps our most surprising finding is that when we scrutinize the flat FRW solutions that have been widely studied as string backgrounds, we find that they almost always suffer from an EFT breakdown. More specifically, we find that in any flat FRW solution in the string landscape, at sufficiently late times, the species scale behaves as $\Lambda_s \sim a^{-\alpha}$ with $\alpha\geq1$. Unless $\alpha=1$, the inequality $H_i \frac{a_i}{a_f} \ll \Lambda_{s,f}$ is eventually violated, leading to the breakdown of the spacetime description. The only spacetimes that avoid an EFT breakdown are those for which $\alpha=1$, and we narrow down these solutions to the following two cases:
\begin{itemize}
    \item A higher-dimensional supersymmetric theory compactified on a negatively curved internal geometry in the pure decompactification limit.
    \item A higher-dimensional supersymmetric theory compactified on a single extra dimension in the pure decompactification limit.
\end{itemize}

Therefore, we find that spatially flat FRW solutions in the string landscape that do not fall into the categories discussed above are unstable, as perturbations inevitably lead to the breakdown of the gravitational effective action. In particular, this applies to any infinite distance limit in which a critical string becomes weakly coupled asymptotically. In such limits, we show that any perturbation around spatially flat FRW backgrounds leads to a breakdown of the $\alpha'$ expansion in the far future.

This instability does not imply that such regions of the string landscape must be discarded. First, the unperturbed background remains a valid string background and provides a meaningful setting for testing Swampland conjectures. Second, we allude to a speculative possibility in Sec.~\ref{SR} that flat backgrounds which break EFT might still admit a non-perturbative consistent stringy descriptions. 

However, the only robust and definitive method to access all infinite distance limits without a breakdown of EFT is introducing negative spatial curvature, which tempers the rate at which the quantum gravity cutoff descends. In general, the scale factor of all expanding solutions in the landscape have leading late-time power-law behavior $a(t)\sim t^p$ where $p<1$ \cite{Rudelius:2021azq} for flat universes and $p=1$ for open universes \cite{Andriot:2023wvg}. Therefore, power-law acceleration cannot be achieved at sufficiently late times \cite{Bedroya:2022tbh}. Using our finding about the necessity of a sufficiently large negative spatial curvature, we place upper bounds on the duration of power-law accelerated expansion at finite times. Since accelerated expansion dilutes spatial curvature, a prolonged period of acceleration would result in a decelerating phase that remains nearly spatially flat for too long, ultimately leading to an EFT breakdown. Specifically, we find that in a cosmological scenario where a quasi-de Sitter phase transitions smoothly into a decelerating expansion, the duration of the quasi-de Sitter phase is constrained by  
\begin{align}
\tau < \frac{1-p}{p(\alpha-1)}\frac{1}{H} \ln\left(\frac{\Lambda_s}{H}\right),
\end{align}  
where \( p \leq 1 \) and \( \alpha \geq 1 \) are determined by the infinite distance limit which the scalar field runs to in the future. In particular, for a spatially flat FRW solution in this direction, we have \( a(t) \sim t^p \) and \( \Lambda_s \sim a^{-\alpha} \).  

Remarkably, this result is proportional to the Trans-Planckian Censorship Conjecture (TCC) bound \cite{Bedroya:2019snp}, which is not a field-theoretic constraint. Our findings here provide an independent argument for TCC in the interior of moduli space that is different from previous works \cite{vandeHeisteeg:2023uxj,Bedroya:2024zta}. In particular, our work reinforces the conclusions of \cite{vandeHeisteeg:2023uxj}, demonstrating that the relationship between the species scale and the scalar potential imposes stringent constraints on the latter.

Finally, we use the ideas developed in this work to study when the inequality
\begin{align}
\left|\frac{\nabla V}{V}\right| \geq \frac{2}{\sqrt{d-2}}\,,
\end{align}
can be saturated in the weakly coupled corners of string theory. We show that Swampland ideas serve as powerful organizing principles, precisely identifying the corners of the string landscape that saturate this bound---namely, the supercritical string theories and decompactifications to an infinite number of dimensions, neither of which belongs to the unitary string landscape.

\section{A Toolbox for the Weakly Coupled String Landscape} \label{sec:2}

In this section, we give a brief review of some of the universal features of the weakly coupled corners of the string theory landscape with positive scalar potential. Many of these features that are known to be true in the string landscape are conjectured to hold for every theory of quantum gravity in the context of the Swampland program \cite{Agmon:2022thq}. Since we focus on the landscape with strictly positive scalar potential, we exclude 11-dimensional M-theory, which  does not have light scalar fields or a scalar potential. 

\subsection{Weak Coupling Limits}

In weakly coupled limits, the scalar potential $V(\phi^I)$ can depend on all the light scalar fields $\{\phi^I|I=1,\hdots,N\}$, some of which control couplings of different sectors in the theory. The scalar field space is equipped with a canonical metric $G_{IJ}(\phi)$ imposed by the kinetic term $\frac{1}{2}G_{IJ}(\phi)\partial_\mu\phi^I\partial^\mu\phi^J$ in the action. This canonical metric on the field space allows us to define quantities such as $|\nabla V|=\sqrt{G^{IJ}\partial_I V\cdot\partial_J V}$ that are invariant under scalar field redefinitions. Throughout this work, we express all dimensionful quantities in reduced Planck scale units. 

 There are two types of asymptotic limits in scalar field space at infinite geodesic distance that contain gravitons and are weakly coupled \cite{Lee:2019wij}:
    
  \begin{itemize}
  \item {\it Decompactification limit}: The size of an internal dimension grows,  leading to a tower of light Kaluza-Klein states whose mass generically scales like the inverse of the diameter of the extra dimension\footnote{There are exceptions to this scaling when there are warping or nontrivial scalar field profiles in the extra dimension\cite{Etheredge:2023odp}.} $m_{\rm KK}\sim L^{-1}$. 
  \item {\it String limit}: The fundamental string that has a graviton as a massless excitation becomes tensionless, and, therefore, there is a tower of weakly coupled string excitations.
  \end{itemize}
  In both cases, there is a tower of particles whose mass falls off exponentially fast in terms of the geodesic distance traveled in the scalar field space parametrized by $\phi$\cite{Ooguri:2006in}.
  \begin{align}
      m(\phi)\sim e^{-c\phi}\,.
  \end{align}
  More specifically, it is known that the ``fall off" of $m$ (defined as $\frac{|\nabla m|}{m})$ is bounded below as
  \begin{align}
      \left|\frac{\nabla m}{m}\right|\geq \frac{1}{\sqrt{d-2}}\,,
  \end{align}
  which was observed in \cite{Etheredge:2022opl} and explained in \cite{Agmon:2022thq}.
  
  \subsection{Scalar Potential}
  
  In all weak coupling limits, the scalar potential falls off exponentially with geodesic distance (parameterized by $\phi$) as \cite{Obied:2018sgi}
   \begin{align}
      V(\phi)\sim e^{-\lambda\phi}\,.
  \end{align}
  The fall off  is observed to be bounded below by $\lambda\geq \frac{2}{\sqrt{d-2}}$. This bound follows from the trans-Planckian Censorship conjecture (TCC)\cite{Bedroya:2019snp}, which postulates that the duration of accelerated expansion in quantum gravity never expands Planckian fluctuations to superhorizon scales, and has been rigorously tested in the weak coupling limits (see \cite{Andriot:2020lea,Andriot:2022xjh,Shiu:2023nph,Shiu:2023fhb} for some examples). The bound
    \begin{align}
      \left|\frac{\nabla V}{V}\right|\geq \frac{2}{\sqrt{d-2}}\,,
  \end{align} 
  has been proven using holographic principle in the infinite distance limits \cite{Bedroya:2022tbh}.
  
\subsection{FRW Solutions at Future Infinity}

 Due to the exponential fall-off behavior of the scalar potential in the weakly coupled corners of scalar field space, no quasi-de Sitter solution in the string landscape can be eternally stable, and it will eventually quantum mechanically tunnel or classically evolve into the asymptotic region of the scalar field space where the scalar potential has an exponential profile. In such corners, the FRW solutions are power-law: 
 \begin{align}
\lim _{t\rightarrow\infty}\frac{a(t)}{t^p}=A\,,     
 \end{align}
 where $a(t)$ is the scale factor, $t$ is the proper time for a stationary comoving observer, and $A,p$ are positive constants. The constants $A$ and $p$ are determined by the factor $\lambda$ in the exponent of the scalar potential and by the sign of the spatial curvature. In the following, we review these attractor solutions that were studied in detail in \cite{Andriot:2023wvg}. Let us first consider non-zero spatial curvature $K=\pm1$. Consider a scalar field $\phi$ with an exponential potential
\begin{equation}
V(\phi) = V_0 e^{-\lambda \phi}\,.
\end{equation}
The scalar equation of motion is
\begin{equation}\label{eoms}
\ddot{\phi} + (d-1)\frac{p}{t}\dot{\phi} - \lambda V_0 e^{-\lambda \phi} = 0\,.
\end{equation}
We assume a solution of the form 
\begin{equation}
\phi(t) = \alpha \ln t + C\,,
\end{equation}
where $C$ is a constant so that $\dot{\phi} = \alpha/t$ and $\ddot{\phi} = -\alpha/t^2$. Matching powers of $t$ in the equation \eqref{eoms} forces $\alpha \lambda = 2$, which implies $\alpha = 2/\lambda$. The overall constant then satisfies
\begin{equation}
V_0 = 2 e^{\lambda C}\frac{(d-1)p - 1 }{\lambda^2}.
\end{equation}
Next, the $d$-dimensional Friedmann equation is
\begin{align}
	\frac{(d-1)(d-2)}{2}\left(H^2 + \frac{K}{a^2}\right)= \frac{1}{2}\dot{\phi}^2 + V(\phi)\,.
\end{align}
Note that $H^2 = p^2/t^2$, $\dot{\phi}^2 = \alpha^2/t^2$, and $V(\phi)\propto 1/t^2$, since $\alpha \lambda = 2$. The curvature term scales as $K/a^2\propto 1/t^{2p}$. For all terms to match a common power of $t$, we must set $p=1$. Hence, we find
\begin{equation}
a(t) = A\hskip 1pt t\,,\quad \phi(t) = \frac{2}{\lambda}\ln t + C\,,
\end{equation}
and the Friedmann equation fixes the combination
\begin{equation}
1- \frac{4}{(d-2)\lambda^2} = -\frac{K}{A^2}\,.
\end{equation}
Note that in the above equation, $K=+1$ implies $\lambda<2/\sqrt{d-2}$ and $K=-1$ implies $\lambda>2/\sqrt{d-2}$. Given that the scalar potentials in the string landscape satisfy $\lambda\geq 2/\sqrt{d-2}$, we conclude that $K=-1$. In general, one can consider FRW cosmologies with any sign for $K$ for any $\lambda$. However, what imposes the correlation between the signs of $K$ and $\lambda-2/\sqrt{d-2}$ is the assumption that the FRW universe has a future infinity and does not end up in a crunch. One can see a summary of different cases in the following table.

\vspace{10pt}

\begin{figure}[H]
    \centering
    \includegraphics[width=\linewidth]{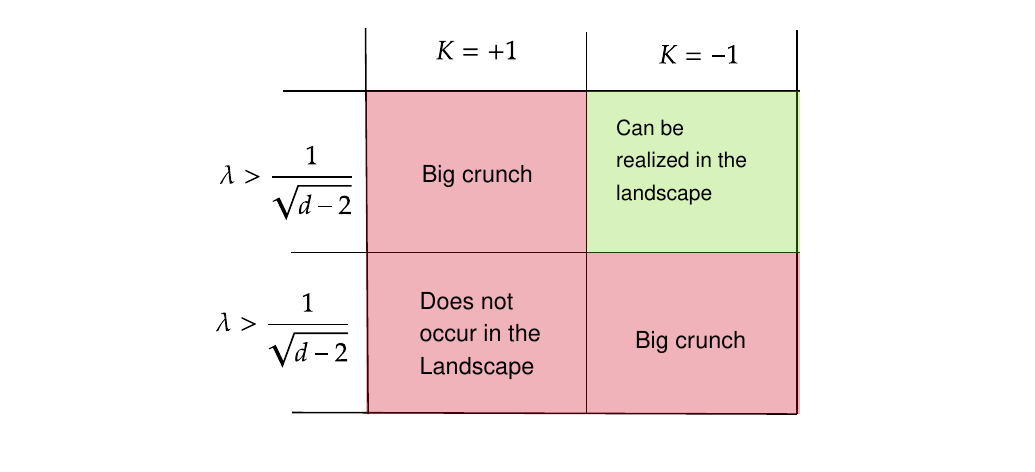}
\end{figure}


In the absence of spatial curvature, we have two regimes:
\begin{align}    
    &\lambda<2\sqrt\frac{d-1}{d-2}\quad \Rightarrow\quad  p=\frac{4}{(d-2)\lambda^2}\,,\nonumber\\
    &\lambda\geq2\sqrt\frac{d-1}{d-2}\quad \Rightarrow\quad  p=\frac{1}{d-1}\,.
\end{align}
In the second case, the scalar potential falls off so rapidly that the kinetic energy of the scalar field dominates the energy density, and the evolution is given by kination. 

\subsection{Species Scale}

The quantum gravity cut-off, also known as the species scale \cite{Dvali:2007hz,Dvali:2012uq}, controls the higher derivative gravitational corrections \cite{vandeHeisteeg:2022btw}. Schematically, the coefficient of a typical $\mathcal{R}^n$ term which is a contraction of $n$ Riemann tensors, has a contribution of the order of $\mathcal{O}(M_{\rm pl}^{d-2}/\Lambda_s^{2(n-1)})$,  which implies a breakdown of any field theory description at energy scales greater than $\Lambda_s$.\footnote{For subtleties regarding the energy dependence of the coefficients of the effective action, see \cite{Bedroya:2024uva,Calderon-Infante:2025ldq,Castellano:2025ljk}.} 

The species scale is a function of the scalar fields and falls off exponentially fast in the weak coupling limit. In the presence of a weakly coupled fundamental string, the species scale is proportional to the string mass, while, in M-theory corners, it is proportional to the higher dimensional Planck mass. In both limits, the species scale decays exponentially fast. The fall off of the species scale is bounded from below and above in weak-coupling limits \cite{vandeHeisteeg:2023uxj}, 
\begin{align}
    \frac{1}{\sqrt{(d-1)(d-2)}}\leq\left|\frac{\nabla\Lambda_s}{\Lambda_s}\right|\leq\frac{1}{\sqrt{d-2}}\,.
\end{align}
The lower bound is saturated by decompactification of exactly one extra dimension, and the upper bound is saturated in the limit that the fundamental string coupling approaches zero while keeping the size of internal dimensions fixed in string units. 

\subsection{Relations Between $V$, $m$, and $\Lambda_s$}

As we explained above, the scalar potential $V$, the mass of the lightest tower $m$, and the species scale $\Lambda_s$ all fall off exponentially fast in weak coupling limits. In fact, there are some relations and inequalities that these quantities satisfy. 

The scaling of $m$ and $\Lambda_s$ is determined in both cases by the nature of the weak coupling limit (i.e.~decompactification or string limit). Therefore, it is natural to expect a relation between the way they fall off. 
 In the weak coupling limits in a string landscape, the result is
 \cite{Castellano:2023stg,Castellano:2023jjt,Etheredge:2024tok}
\begin{align}\label{IDI}
    \frac{\nabla m}{m}\cdot\frac{\nabla \Lambda_s}{\Lambda_s}=\frac{1}{d-2}\,,
\end{align}
This identity is not expected to hold everywhere in the string landscape but rather, is conjectured to be replaced with the inequality $\frac{\nabla m}{m}\cdot\frac{\nabla \Lambda_s}{\Lambda_s}\leq\frac{1}{d-2}$ \cite{Bedroya:2024uva}.

In any FRW solution, we expect the Hubble parameter to be smaller than both the species scale \cite{vandeHeisteeg:2023uxj} and the mass scale of the tower \cite{Rudelius:2022gbz}. These two conditions ensure that the field theory description of the spacetime does not breakdown due to higher-order corrections or a tower of states. If the Hubble parameter exceeds the string mass, field theory breaks down, while if it exceeds the KK mass, we must use the higher-dimensional field theory, since the size of the internal dimensions is then greater than the observable universe. 
In fact, the mass scale of the lightest tower of states is always smaller than the species scale. Moreover, if the energy density is not dominated by the kinetic energy of the scalar field, we have $V\sim H^2$ in Planck units. 
We therefore find
\begin{align}\label{UI2}
    \sqrt{V}\leq m\leq \Lambda_s\,,
\end{align}
which we will use in the next section to demonstrate the breakdown of EFT in some spatially flat FRW spacetimes. Note that some of the states in the tower have spin-2 and the bound $H\lesssim m$ is similar to the Higuchi bound \cite{Higuchi:1986py} but follows from a different line of reasoning.

\section{A Novel Mechanism for Breakdown of EFT}\label{sec:3}

As we observed in the previous section, the future behavior of any FRW solution with a positive scalar potential in the string landscape is characterized by a power-law FRW solution, $\lim_{t\rightarrow\infty} \frac{a(t)}{t^p} = A$, driven by an exponential scalar potential, $V(\phi)\propto e^{-\lambda \phi}$. Furthermore, the quantum gravity cutoff, $\Lambda_s$, and the mass of the lightest tower of weakly coupled states also decay exponentially with the scalar field strength. With these observations in mind, we are ready to formulate a new mechanism for the breakdown of EFT in quantum gravity.

\subsection{The mechanism}

Consider a Hubble-sized scalar perturbation in this background. For the field theory to describe an almost homogeneous FRW solution, we require the quantum gravity cutoff to remain significantly greater than the proper momentum of this perturbation. Thus, the following inequality must hold: \begin{align}\label{ME} H_i \frac{a_i}{a_f} \ll \Lambda_{s,f} \,. \end{align} If this inequality is violated, the higher-derivative corrections in the action become a divergent series, and the metric description of spacetime breaks down. Therefore, to ensure that the gravitational effective field theory (EFT) does not break down at future infinity, the spacetime must satisfy the inequality \eqref{ME}. Note that we consider a Hubble-sized perturbation to ensure that its evolution is well approximated by linearized equations in an FRW background. Although an almost FRW background may contain strong inhomogeneities on smaller scales that collapse into gravitational structures, strong nonlinearity on the Hubble scale would prevent approximating the spacetime with an FRW solution.

\subsection{Distinction from Trans-Planckian Issues}\label{sec:5}

The inequality \eqref{ME} is similar in nature to the Trans-Planckian Censorship Conjecture (TCC), since both require that Hubble-sized perturbations must not exceed the quantum gravity cutoff. However, our condition is imposed in the future time direction rather than the past, which we argue leads to a significant difference. In the future direction, perturbations exceeding the cutoff result in a breakdown of effective field theory (EFT), whereas in the past, this does not necessarily occur. This distinction explains why TCC is a quantum gravity constraint rather than a fundamental field theory constraint. To illustrate this difference, it is instructive to examine why a violation of TCC does not necessarily imply a breakdown of EFT. It is often stated that the trans-Planckian issue is not a genuine field theory breakdown since the modes that emerge from the Planckian regime can be taken to be in the vacuum state. However, this tautological solution fails when we consider any state other than vacuum, because that means at some point in the past, a state that emerged the trans-Planckian regime was not exactly in the vacuum. We argue here that the actual resolution is more nuanced.

\subsubsection{Black Holes}

Before addressing the avoidance of trans-Planckian problems in cosmological settings, it is useful to analyze this issue in the context of black holes. The main concern regarding non-vacuum states in setups that might seemingly exhibit trans-Planckian issues is that such states cannot be described using EFT and might also generate trans-Planckian energy densities. While it is true that near the horizon of black holes, the stationary observer's EFT breaks down due to Planckian Unruh radiation, a free-falling observer does not experience such a breakdown of EFT or encounter trans-Planckian energy densities in the Unruh vacuum \cite{Unruh:1976db}. In other words, performing quantization in the free-falling observer's frame reveals a nontrivial Hilbert space containing many states in which the energy-momentum tensor does not have Planckian expectation values. These states remain well-described by EFT in the free-falling frame and project onto mixed states outside the horizon, encoding Hawking radiation.

Resolving the trans-Planckian problem is more challenging in expanding spacetimes than in black holes. As highlighted in \cite{Bedroya:2024zta}, unlike black holes, the trans-Planckian issue in expanding spacetimes is observable by a free-falling observer. However, in de Sitter spacetime, a resolution analogous to the black hole case exists.  

\subsubsection{de Sitter Space}

Consider two coordinate systems in de Sitter spacetime: the flat slicing and the global slicing. The metric in flat slicing coordinates is given by
\begin{align}
    \dd s^2 = \frac{1}{H^2\eta^2}(-\dd \eta^2 + \dd \vec{x}^2),
\end{align}
and in global coordinates, it is given by
\begin{align}
    \dd s^2 = \frac{1}{H^2\cos^2\tau}\left(-\dd \tau^2 + \dd \Omega_{d-1}^2\right).
\end{align}
For a free massive scalar field with mass $m$, the solutions in flat coordinates take the form $\phi_F(k,\eta)e^{i\vec{k}\cdot\vec{x}}$, where $\phi_F(k,\eta)$ satisfies
\begin{align}
       \phi_F'' - \frac{d-2}{\eta}\phi_F' + \left(k^2 + \frac{m^2}{H^2\eta^2}\right)\phi_F = 0.
\end{align}
The general solution is given by
\begin{align}\label{FS}
    \phi_{F}(k,\eta) = i\sqrt{\frac{\pi}{4H}}(-H\eta)^{\frac{d-1}{2}} H^{(1)}_{i\nu}(-k\eta),
\end{align}
where $i\nu = \sqrt{(\frac{d-1}{2})^2 - (\frac{m}{H})^2}$ and $H^{(1)}$ is the Hankel function of the first kind. These modes correspond to positive frequency modes in the past ($\eta\to-\infty$). 

In global coordinates, the constant-time slices are compact, leading to a discrete spectrum for the Laplacian. In $d$ spacetime dimensions, the eigenfunctions of the Laplacian are the well-known spherical harmonics, denoted as $Y_k(\Omega)$, with $k^2 = L(L + d - 1)$ and $L \in \mathbb{Z}_+$. The spherical harmonics contain additional indices, but we track only $k$ due to its role in the time evolution of modes. The solutions take the form $\phi_G(L,\tau) Y_k(\Omega_{d-1})$, where $\phi_G(L,\tau)$ satisfies
\begin{align}
    \phi_G'' - \frac{(d-2)\sin\tau}{\cos\tau}\phi_G' + \frac{L(L + d - 1) + m^2}{H^2\cos^2\tau}\phi_G = 0.
\end{align} 
The general solution is
\begin{align}\label{GS}
    \phi_{G}(L,\tau) &= \frac{\cos^\Delta\tau(\theta(\tau) - \theta(-\tau))}{\sqrt{d-1 - 2\Delta}} 
    \times {}_2F_1\bigg[ \frac{\Delta+L}{2}, \frac{\Delta-d+2-L}{2}, \frac{3}{2} + \Delta - \frac{d}{2} \Big| \cos^2\tau \bigg],
\end{align}
where $\Delta = \frac{d-1}{2} + \sqrt{(\frac{d-1}{2})^2 - (\frac{m}{H})^2}$ and ${}_2F_1$ is the hypergeometric function. 

The Bunch-Davies vacuum is defined as the state annihilated by the corresponding annihilation operators in either flat or global coordinates \cite{Loparco:2023akg, Mottola:1984ar}. However, while flat coordinates yield a continuous spectrum of annihilation operators, global coordinates yield a discrete spectrum. This results in a discrepancy between the Fock spaces of the two coordinate systems. 

To illustrate this difference concretely, consider the energy-momentum tensor for eigenstates of the particle number operators $a_k^\dagger a_k$, where $a_k$ is the annihilation operator associated with the mode of comoving momentum $k$. The energy-momentum tensor for a free massive scalar field is given by
\begin{align}
    T_{\mu\nu} = \partial_\mu\phi\partial_\nu\phi - \frac{g_{\mu\nu}}{2} (\partial_\rho\phi\partial^\rho\phi + m^2\phi^2)\,.
\end{align}
The $T^0_0$ component evaluated on an eigenstate of particle numbers is given by 
\begin{align}
    T^{0}_0&=\int d^{d-1}k\,(H\eta)^2 n_k \Big(|\phi'_{F}(k,\eta)|^2-((kH\eta)^2+m^2)|\phi_{F}(k,\eta)|^2\Big)\,,
\end{align}
where $n_k$ is the particle number density in  momentum space. 
We have used normal ordering to get a vanishing energy-momentum tensor on the Bunch-Davies vacuum. 
To ensure that the energy-momentum tensor is UV-finite, we only include states with $n_k=0$ above a UV cut-off. 

Note that the  integral above is divergent on the 
$\eta\rightarrow -\infty$ surface for nonzero particle numbers due to the  behavior of Hankel functions such as $H_{i\nu}^{(1)}(-k\eta)\sim (-k\eta)^{-1/2}$ in the limit of large argument.
However, if we consider one-particle states obtained by quantization in  global coordinates, we do not encounter this singularity since those solutions \eqref{GS} are regular in the spacetime. 
Note that, by choosing different modes, we impose different boundary conditions. 
For instance, if we express the solutions $\phi_GY_k(\Omega_{d-1})$, where $\phi_G$ is defined in \eqref{GS} in terms of  flat coordinates, we find that 
\begin{align}\label{AB}
   \phi\sim r^{-2\Delta}\,,
\end{align}
at fixed $\eta$ and large $r$,
where $r$ is the radius in  flat slicing coordinates. This is because the coordinate $\tau$ is related to $\eta$ and $r$ via
\begin{align}
    \tau=2\tan^{-1}\left(\frac{\sqrt{1+X^2}-1}{X}\right)\,,
\end{align}
where
\begin{align}
    X=\frac{(H\eta)^{-1}-H\eta}{2}+\frac{H^2r^2}{2H\eta}\,.
\end{align}
For a fixed $\eta$, large $r$ corresponds to large $X$ and correspondingly $\tau\simeq \frac{\pi}{2}$. At large radii, $Hr\propto \sqrt{X}$ and $\tau\simeq \pi/2-1/X$ leading to $ \pi/2-\tau\propto 1/r^2$. On the other hand, the hypergeometric function \eqref{GS} decays as $(\pi/2-\tau)^\Delta$, which leads to the asymptotic behavior \eqref{AB}. 
This boundary condition differs from  the one followed by the solutions in \eqref{FS}.

\begin{figure}[t!]
    \centering
    \includegraphics[width=\linewidth]{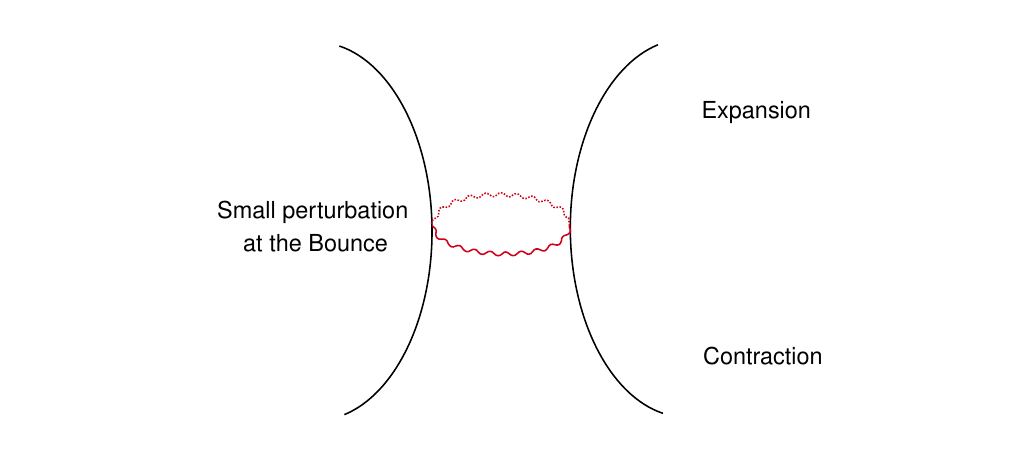}
\caption{Sufficiently small perturbation at the throat of global de Sitter space will evolve smoothly in the future and past directions.}
\label{SP}
\end{figure}

We conclude that if one considers small-energy perturbations at the throat in the global slicing, there will be no singularity in the backreacted geometry. 
We note that the states that we consider must have sufficiently small energies at time $\tau=0$, to ensure that the backreaction remains small (see Figure~\ref{SP}). Since the isometries of de Sitter allow us to move every point on this slice, we conclude that every observer in de Sitter has access to local states that will not create a singularity in the spacetime.

Importantly, the smooth perturbations in the global slicing take a very specific form when expanded in terms of the flat slicing modes \eqref{FS}. Since the smooth perturbations decay as $r^{-2\Delta}$ at large $r$, that means when they are expanded in terms of the flat slicing modes \eqref{FS} using Fourier transform, the coefficient $\phi_k$ of the mode with comoving momentum $k$, has the following scaling law at small $k$,
\begin{align}
    \phi_k\sim k^{2\Delta-(d-1)}\,.
\end{align}
The above equation shows that avoiding a singularity in the past, depends on the the low-momentum behavior of the perturbation, since, as we move in the past direction, the long-range perturbation shrinks and becomes relevant. We will later see that the above idea can be generalized to deduce that no accelerated expansion suffers from a field theoretic trans-Planckian issue.

\begin{figure}[t!]
    \centering
    \includegraphics[width=\linewidth]{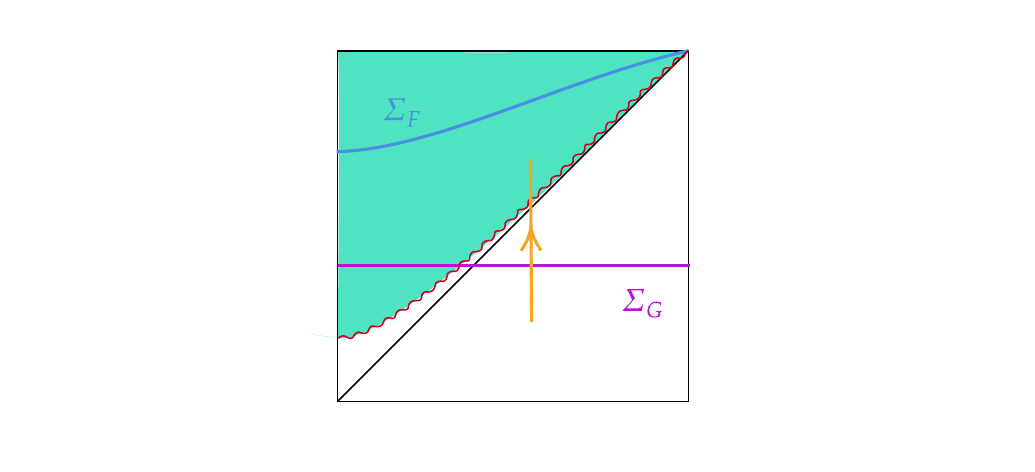}
    \caption{The shaded region in  de Sitter space is parameterized by the flat coordinates. A generic state in the Fock space defined on the hypersurface $\Sigma_F$ traces back to a singularity in the past illustrated by the red squiggly line. However, the Fock states for the free-falling observer (orange line) will result in a smooth spacetime. These states correspond to the Fock states defined in  global coordinates on the hypersurface $\Sigma_G$.}
    \label{dS}
\end{figure}

This is reminiscent of what happens for black holes. The parallel becomes even clearer when we recall that the comoving observer in the global slicing is a free-falling observer that smoothly traverses the $\eta=-\infty$ null surface at finite proper time (see figure \ref{dS}). 
Therefore, just as the free-falling observer passing through the event horizon of the black hole observes a well-behaved Hilbert space, so does the observer free-falling through the null surface at $\eta=-\infty$. 

\subsubsection{Milne space}

A similar resolution to the trans-Planckian issue exists in  Milne space, which is  a reparameterization of  Minkowski space. The metric of Milne space is 
\begin{align}
    \dd s^2=-\dd t^2+t^2(\dd r^2+\sinh^2 r\,\dd \Omega_{d-2}^2)\,.
\end{align}
There are quasi-stationary solutions to the equations of motion given by \cite{Saharian:2020uiu}
\begin{align}
    \phi_\gamma=t^\frac{2-d}{2}H^{(1)}_{i\sqrt{\gamma^2-(\frac{d-2}{2})^2}}(mt)\,,
\end{align}
where $\gamma$ is set by the eigenvalue of the Laplacian on the hyperbolic space. Similar to the de Sitter modes, these modes diverge at the past surface $t=0$. However, since the Milne spacetime is a subspace of  Minkowski spacetime, we know that there are nontrivial states in the Hilbert space that will necessarily not create a singularity in the past at $t=0$. 

\begin{figure}[t!]
    \centering
    \includegraphics[width=\linewidth]{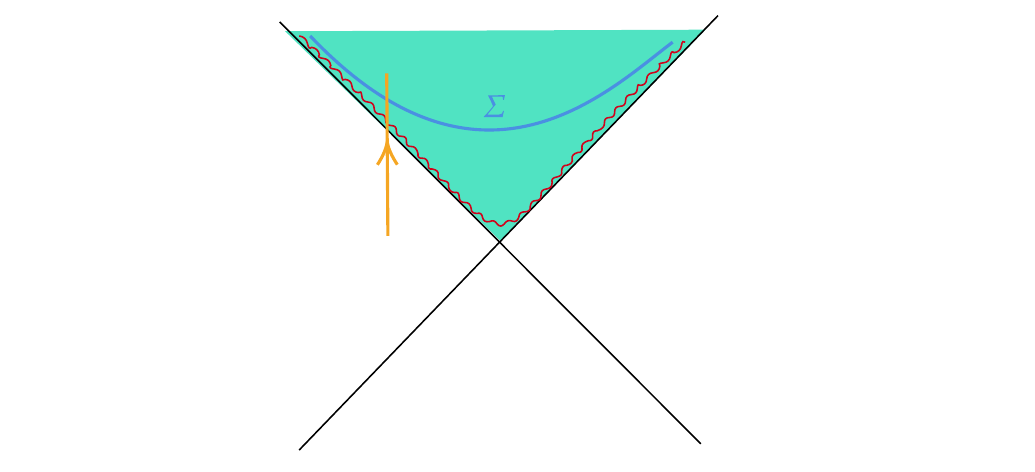}
    \caption{The shaded region in  Minkowski space is parameterized by the Milne coordinates. A generic state in the Fock space defined on the hypersurface $\Sigma$ traces back to a singularity in the past illustrated by the red squiggly line. However, the Fock states for the free-falling observer (orange line) will result in a smooth spacetime.}
    \label{Milne}
\end{figure}

Therefore, thanks to the existence of a free-falling observer who can cross the past null surface $t=0$, as illustrated by the orange arrow in Figure \ref{Milne}, the quantum states in that observer's rest frame will provide a Hilbert space of well-behaved states. 

\subsubsection{Accelerated Expansion}

In the following, we use our observation of the important role played by the long-range behavior of the perturbation in smoothening of the past, to show that, in any power-law accelerated expansion, there is no actual breakdown of EFT. Consider a spacetime which would naively create trans-Planckian issue due to an expansion from time $t_1$ to $t_2$ that stretch Planckian modes to Hubble sized modes. We show that there are Hubble sized perturbation at time $t_2$ that are indeed smooth at $t=t_1$. More specifically, we show that the breakdown of EFT is caused by a bad choice of mode expansion and nothing more. 

The key observation is that, if one considers the right superposition of super-horizon modes at $t=t_1$, they will become irrelevant at the later time due to the suppression of the amplitudes of the frozen modes by the Hubble friction. However, the addition of these almost constant modes to an otherwise UV perturbation ensures that $|\frac{\nabla\phi}{\phi}|\ll M_{\rm pl}$ is small and well-within the field theory approximation. Note that here $\nabla$ refers to gradient with respect to the spacetime metric.

Suppose we have a power-law acceleration $a(t)\propto t^p$ where $p>1$ from $t=t_1$ to $t=t_2$ such that  
\begin{align}
    \frac{a(t_2)}{a(t_1)}l_{\rm pl}=\frac{M_{\rm pl}}{H(t_2)}\,.
\end{align}
We will find a solution to the equation of motion of a massless scalar field such that $|\frac{\nabla\phi}{\phi}|$ remains much smaller than the Planck scale throughout this time interval and is of the order of the Hubble parameter at time $t=t_2$. In the conformal coordinates, the spacetime approaches 
\begin{align}
    \dd s^2=a(\eta)^2(-\dd \eta^2+\dd \vec x^2)\,,
\end{align}
where $\eta\propto-t^{1-p}<0$ and
\begin{align}
    a(\eta)\propto (-\eta)^{-\frac{p}{p-1}}\,.
\end{align}
We have neglected the effect of the spatial curvature due to the fact that accelerated expansion dilutes spatial curvature contribution to the critical energy density.

The equation of motion for a massless scalar field with comoving momentum $k$ takes the form
\begin{align}
    \phi_k''+(d-1)\frac{a'}{a}\phi_k'-k^2\phi_k=0\,.
\end{align}
The prime  represents derivative with respect to the conformal time $\eta$. In the spacetime of our interest, the above equation takes the following form
\begin{align}\label{eom}
    \phi_k''-\frac{(d-1)p}{(p-1)\eta}\phi_k'-k^2\phi_k=0\,.
\end{align}
If $k\gg \frac{(d-1)p}{(p-1)\eta}$, the friction term is negligible and the norm of $\phi_k$ remains constant. However, when $k\ll\frac{(d-1)p}{(p-1)\eta} $, the mode exits the Hubble horizon, and the Hubble friction leads to a decaying solution. Therefore, for such values of $k$, there exists a power-law solution 
\begin{align}\label{scl}
    \phi_k\propto (-\eta)^{1+\frac{(d-1)p}{(p-1)}}\propto a^{-(d-1)-\frac{p-1}{p}} \,.
\end{align}
The above solution follows from the balancing  the first two terms in the equation of motion \eqref{eom}. Note that the first two terms also have a constant solution, but that is no longer a solution when we include the third term unless $k=0$. Nonetheless, the differential equation \eqref{eom} has a second solution which we will not use to construct our smooth modes.

We only consider modes that have sub-horizon proper momenta at $t=t_1$. This corresponds to modes with comoving momenta $k\geq k_{\rm min}=H(t_1)a(t_1)$. For such modes, we can relate their size at time $t_2$ to their size at time $t_1$ by first finding when they exit the Hubble horizon and then applying the scaling law \eqref{scl} to them. A mode with comoving momentum $k$ becomes Hubble sized at scale factor $a_k$ and Hubble parameter $H_k$ such that
\begin{align}
    k\sim a_k H_k\sim a_k^{1-p^{-1}}\quad\Rightarrow\quad a_k\propto k^\frac{p}{p-1}\,.
\end{align}
Suppose $k_{\rm max}=H(t_2)a(t_2)$ is the comoving momentum of the mode that becomes Hubble sized at $t=t_2$.
\begin{align}
    a_k\simeq \left(\frac{k}{k_{\rm max}}\right)^\frac{p}{p-1}a(t_2)\,.
\end{align}
Now we can find the diluted amplitude at $t=t_2$ using the scaling law \eqref{scl} as follows
\begin{align}
    |\phi_k(t_2)|&\simeq |\phi_k(t_1)|\left(\frac{a(t_2)}{a_k}\right)^{-(d-1)-\frac{p-1}{p}}=|\phi_k(t_1)|\left(\frac{k}{k_{\rm max}}\right)^{\frac{p(d-1)}{p-1}+1}.
\end{align}
Suppose we consider a mode with a power-law Fourier profile $|\phi_k(t_1)|\propto k^{-\beta}$ for $k\in[k_{\min},k_{\max}]$. Then we have
\begin{align}
    \left|\frac{\nabla \phi}{\phi}\right|
    &\sim a^{-1}(t)\frac{\int_{k_{\min}}^{k_{\max}} k^{d-1}\,\dd k|\phi_k(t)|}{\int_{k_{\min}}^{k_{\max}} k^{d-2}\,\dd k|\phi_k(t)|}\nonumber\\
    &=a^{-1}(t)\frac{\int_{k_{\min}}^{a(t)H(t)}k^{d-1}|\phi_k(t)|\,\dd k+\int_{a(t)H(t)}^{k_{\max}}k^{d-1}|\phi_k(t)|\,\dd k}{\int_{k_{\min}}^{a(t)H(t)}k^{d-2}|\phi_k(t)|\,\dd k+\int_{a(t)H(t)}^{k_{\max}}k^{d-2}|\phi_k(t)|\,\dd k}\nonumber\\
    &=a^{-1}(t)\frac{\int_{k_{\min}}^{a(t)H(t)}k^{d-1-\beta}(\frac{k}{a(t)H(t)})^{\frac{p(d-1)}{p-1}+1}\,\dd k+\int_{a(t)H(t)}^{k_{\max}}k^{d-1-\beta}\,\dd k}{\int_{k_{\min}}^{a(t)H(t)}k^{d-2-\beta}(\frac{k}{a(t)H(t)})^{\frac{p(d-1)}{p-1}+1}\,\dd k+\int_{a(t)H(t)}^{k_{\max}}k^{d-2-\beta}\,\dd k}\,.
\end{align}
Note that here $\nabla$ continues to refer to the gradient with respect to the spacetime metric. In the second line, we have estimated the fraction assuming no finely-tuned cancellation between different modes. The first terms dominate at $t=t_2$ and the second terms dominate at $t=t_1$. If we choose $\beta<d-1+\frac{p(d-1)}{p-1}$, that ensures that, at time $t=t_1$, $|\frac{\nabla \phi}{\phi}|$ is dominated by the large $k$ limit of the integral and therefore we have $|\frac{\nabla \phi}{\phi}|(t_2)\sim H(t_2)$. However, if we choose $\beta>d-1$, the second integral will be dominated by its IR behavior and therefore, it will not be proportional to $k_{\max}$, but to $k_{\min}$. The integral is then 
\begin{align}
    \left|\frac{\nabla \phi(t_1)}{\phi(t_1)}\right|\sim k_{\min}a(t_1)=\frac{k_{\min}}{k_{\max}}M_{\rm pl}\ll M_{\rm pl}\,.
\end{align}
Hence we find that the solution avoids Planckian issues thanks to the friction term which suppresses superhorizon modes. Note that the two inequalities 
\begin{align}
    d-1<\beta<d-1+\frac{p(d-1)}{p-1}
\end{align}
can be simultaneously satisfied for every $p>1$. Moreover, note that this resolution only applies in the presence of the friction term, which is why it cannot resolve a naive breakdown of EFT in the future direction as discussed in \eqref{ME} in expanding universes.

\section{EFT Breakdown in the $K=0$ String Landscape}\label{sec:4}

In this section, we show that, in the absence of spatial curvature, most FRW solutions in the string landscape violate the inequality \eqref{ME}. Consider an FRW solution with a canonically normalized scalar field, $\phi$, that runs to infinity and parameterizes the cosmological trajectory in moduli space.
 We will consider different possibilities, including those where the energy density is not dominated by the scalar field. We divide our analysis into three cases: (I) Non-negligible scalar potential ($V \sim \dot{\phi}^2 \sim H^2$); (II) Negligible scalar potential with non-negligible kinetic energy, also known as kination ($\dot{\phi}^2 \sim H^2 \gg V$); and (III) Negligible scalar field kinetic energy ($\dot{\phi}^2 \ll H^2$). We will show that the third case can occur for a finite period of time, but it can not govern the evolution at sufficiently late times.

 We use the symbols $\sim$, $\gtrsim$, and $\lesssim$ to compare the growth rates of expressions at future infinity. For example, for time-dependent positive parameters $U(t)$ and $V(t)$: \begin{align} \label{tilde}
 U \lesssim V\quad  &\leftrightarrow\quad  \limsup_{t\rightarrow\infty} \frac{U(t)}{V(t)} <\infty,\nonumber\\
 U \gtrsim V \quad &\leftrightarrow\quad  \liminf_{t\rightarrow\infty} \frac{U(t)}{V(t)} >0,\nonumber\\
 U\sim V\quad  &\leftrightarrow\quad  0 < \liminf_{t\rightarrow\infty} \frac{U(t)}{V(t)} \leq \limsup_{t\rightarrow\infty} \frac{U(t)}{V(t)} < \infty \,. \end{align}
Note that $U\sim V$ only implies that $U$ and $V$ have similar scaling dependence on time as $t\rightarrow \infty$, but does not imply that $U$ and $V$  are of the same order. For example, we can have $U\sim V$ even in cases where $U/V\gg1$ provided it approaches a constant as $t \rightarrow \infty$. Throughout our analysis in this section, we continue to work in reduced Planck units and assume that the scalar potential takes an exponential form, $V \sim e^{-\lambda \phi}$, which is justified in the infinite distance limits of the moduli space \cite{Obied:2018sgi}.


\subsection{Case I: $V \sim \dot{\phi}^2 \sim H^2$}
From Eq.~\eqref{IDI}, we know that, in the weakly coupled regime, we have $\frac{\nabla m}{m} \cdot \frac{\nabla \Lambda_s}{\Lambda_s} = \frac{1}{d-2}$. Moreover, from the inequality \eqref{UI2}, we know that the validity of EFT requires that, everywhere in the weakly coupled regime, we have $\sqrt{V} \lesssim m$. In particular, along the infinite distance direction of $-\nabla \Lambda_s$, the fall-off of $\sqrt{V}$ must be greater than that of $m$. Therefore, we obtain the following inequality: 
\begin{align}\label{eq1} 
\frac{\nabla \sqrt{V}}{\sqrt{V}} \cdot \frac{\nabla \Lambda_s}{\Lambda_s} \geq \frac{\nabla m}{m} \cdot \frac{\nabla \Lambda_s}{\Lambda_s} = \frac{1}{d-2}\,,
\end{align} 
which holds asymptotically as $\phi \rightarrow \infty$. This inequality is independent of the solution and can be applied to any point in the weakly coupled regime of the scalar field space.

Now suppose the potential in the direction of $\phi$ which parameterizes the trajectory of the scalar field space scales as $V \sim e^{-\lambda \phi}$. In the infinite distance limits where the scalar potential is a non-negligible part of the energy density, this direction is the same as the direction of steepest descent for the scalar field potential. 
If the species scale scales as $e^{-c\phi}$, then the inequality \eqref{eq1} implies
\begin{align}\label{eq2} 
c \geq \frac{2}{(d-2)\lambda} \quad \Rightarrow \quad \Lambda_s \lesssim e^{-\frac{2}{(d-2)\lambda} \phi}\,.
\end{align} 
Note that $\phi$ is not necessarily the direction of steepest descent for $\Lambda_s$, so the above inequality does not imply $|\nabla \Lambda_s / \Lambda_s| \geq \frac{2}{d-2}$. The scale factor increases as $t^p$, where $p = \frac{4}{(d-2)\lambda^2}$. Since $t \sim 1/H \sim 1/\sqrt{V}$, we find that
\begin{align} a \sim e^{\frac{2}{(d-2)\lambda} \phi}\,. \end{align}
By combining \eqref{eq1} and \eqref{eq2}, we find: 
\begin{align}\label{ME2}
a(t) \lesssim \frac{1}{\Lambda_s(t)}\,. 
\end{align} 
Unless this inequality is saturated, we find that $\Lambda_s$ falls off strictly faster than $1/a$.  Consequently, the inequality \eqref{ME} will be violated and the EFT breaks down.

Next, let us examine the condition under which the inequality \eqref{ME2} is saturated. This occurs if the inequality \eqref{eq1} is saturated, meaning that, in the direction of $\nabla \Lambda_s$, the scalar potential scales as $m^2$. If the species scale is given by the string mass for a light fundamental string, this means that the scalar potential scales as $M_s^2 M_{\rm pl}^{d-2}$, which is of order the tree-level closed string contribution. However, this cannot happen in critical string theories due to the conformal symmetry of the string worldsheet.\footnote{In supercritical string theories (e.g.~time-like linear dilaton \cite{Hellerman:2006nx, Aharony:2006ra}) one can in principle obtain such scalar potentials but such theories have not been proven to be unitarity.} On the other hand, if the species scale scales with the higher-dimensional Planck mass, the scalar potential is proportional to $m_{\rm KK}^2$. This limit occurs when the internal geometry has negative curvature of the order of $m_{\rm KK}^2$. 

A family of such examples are superavities compactified on $T^n\times \mathcal{H}$ where $\mathcal{H}$ is a hyperbolic space \cite{DeLuca:2021pej}.
 In such limits, we can also conclude that the Hubble parameter will scale as $\sqrt{V} \sim m_{\rm KK}$. Even though the Hubble scales as $m_{\rm KK}$, it does not mean that they are of the same order (recall \eqref{tilde} and the discussion below). In fact, we can create an arbitrarily large separation of scales between the two to ensure that the cosmology is well-approximated by the lower-dimensional description. The key is to realize that for an exponential potential $V\propto e^{-\lambda\phi}$ there is a family of exact solutions given by $\phi(t)=\frac{2}{\lambda}\ln(t)+C$ where $C$ is an arbitrary constant. The constant $C$ also determines the separation of scales between $H=\frac{4}{(d-2)\lambda^2 t}$ and $m_{\rm KK}\propto e^{-\lambda\phi/2}$. By choosing an arbitrarily negative $C$, we can create an arbitrarily large ratio of $m_{\rm KK}/H$. 

Therefore, we conclude that in any infinite-distance limit where $V \sim \dot{\phi}^2 \sim H^2$ and $m \gg H$, we have
\begin{align} 
\Lambda_s \sim a^{-\alpha}\,, 
\end{align} 
for some $\alpha \geq 1$, where $\alpha=1$ occurs only for infinite-distance limits that correspond to pure decompactification where the theory decompactifies to a higher-dimensional theory compactified on a negatively curved space. 

In decompactification limits, the fact that the infinite-distance limit saturating the inequality $\alpha \geq 1$ must be a pure decompactification limit allows us to conclude that the higher-dimensional theory must be supersymmetric. The key point to see why is that saturating the inequality \eqref{ME2} strongly depends on the chosen direction in field space. Specifically, if we combine the direction in scalar field space corresponding to the decompactification of the internal negatively curved space with other moduli, the inequality would no longer be saturated. However, such mixing is unavoidable if the higher-dimensional theory has a strictly positive scalar potential that depends on some scalar field. This is because the equation of motion for the scalar field drives it along the direction of steepest descent of the scalar potential, which necessarily involves movement in the moduli space of the higher-dimensional theory.

For example, suppose the $D$-dimensional theory has a scalar potential $V \sim e^{-\lambda \phi_1}$ in some weakly coupled regime. After compactifying the theory on a negatively curved space with a canonically normalized volume modulus $\phi_2$, the lower-dimensional scalar potential will have two contributions: one from the internal curvature, proportional to $e^{-2\sqrt\frac{D-2}{(D-d)(d-2)}\phi_2}$, and another proportional to $e^{-\lambda \phi_1}$. Consequently, if the scalar field follows the direction of steepest descent, it will eventually move in a combination of the $\phi_1$ and $\phi_2$ directions, such that the two contributions compete, and this direction is no longer purely a decompactification direction.

Thus, we conclude that the only way to saturate the inequality \eqref{ME2} via a decompactification limit is if the higher-dimensional theory has a vanishing scalar potential, which is expected to occur only in supersymmetric theories. While such constructions may break supersymmetry at energy scales below the Kaluza-Klein (KK) scale, supersymmetry is restored in the decompactification limit, where the higher-dimensional theory is recovered.


\subsection{Case II: $\dot{\phi}^2\sim H^2,\lim_{t\rightarrow\infty}\frac{V}{H^2}\rightarrow 0$}

When the scalar potential is too steep ($\lambda>2\sqrt{\frac{d-1}{d-2}}$), the kinetic energy of the scalar field dominates over the scalar potential and the scalar potential can be neglected. There is a tracking FRW solution for a free scalar field known as the kination solution and is given by
\begin{align}\label{KDS1}
\phi(t)&=\sqrt\frac{d-2}{d-1}\ln(t)+C\,,
\end{align}
for a constant $C$. The scale factor  for this solution is 
\begin{align}\label{KDS2}
a(t)\propto t^\frac{1}{d-1}\,.
\end{align}
Using the equations \eqref{KDS1} and \eqref{KDS2}, we can express the scale factor in terms of the scalar field as
\begin{align}\label{KDS3}
a(t)\sim e^{\frac{1}{\sqrt{(d-1)(d-2)}}\phi}\,.
\end{align}
Using the emergent string conjecture, authors of \cite{vandeHeisteeg:2023uxj} found the following lower bound on the fall-off of the species scale in \textit{any} infinite distance direction and not just the direction of steepest descent,  
\begin{align}\label{KDS4}
\left|\frac{\nabla\Lambda_s}{\Lambda_s}\right|\geq\frac{1}{\sqrt{(d-1)(d-2)}}\,.
\end{align}
 We use the sign convention that $\phi\rightarrow+\infty$ is the infinite distance direction at which $\Lambda_s$ falls off. Therefore, $\nabla\Lambda_s/\Lambda_s$ is negative and we find 
\begin{align}\label{KDS5}
\Lambda_s\lesssim e^{-\frac{1}{\sqrt{(d-1)(d-2)}}\phi}\,.
\end{align}
Authors of \cite{vandeHeisteeg:2023uxj} also identified that the inequality \eqref{KDS4} is saturated if and only if  exactly one dimension decompactifies in the infinite distance direction of $\phi$. Therefore, by combining \eqref{KDS3} and \eqref{KDS5}, we find
\begin{align}\label{ME3}
a(t)\lesssim \frac{1}{\Lambda_s(t)}\,.
\end{align}

In kination cosmologies, the above inequality is saturated if and only if \eqref{KDS5} is saturated. Therefore, (i) if the cosmology is kination and (ii) there is exactly one dimension that decompactifies in the infinite distance limit, our new bound is saturated. Similarly to the spacetime that saturates the inequality \eqref{ME3} in Case I, the ratio of the Hubble scale to the KK scale approaches a constant in the decompactification limit. However, the ratio, $H/m_{\rm KK}\propto e^{\frac{C}{\sqrt{(d-1)(d-2)}}}$, can be arbitrarily small since $C$ can be arbitrarily negative. We conclude that any infinite distance limit that is described by an FRW cosmology satisfies 
\begin{align}
    \Lambda_s\sim a^{-\alpha}\,,
\end{align}
for some $\alpha\geq1$, and the inequality  is saturated ($\alpha=1$) for limits described by a supersymmetric theory compactified on a one-dimensional manifold.   The manifold could be an interval or a circle with branes and warping along the extra dimension. Similar to case I, the supersymmetry of the limiting theory is required to ensure that there are no additional moduli in the higher dimensional theory that can decrease the scalar potential faster. In other words, this condition makes sure we have indeed tracked the steepest descent of the scalar potential to arrive at this infinite distance limit. 

To understand the uplift of the spacetime that saturates \eqref{ME3} to the higher-dimensional space, consider the $(d+1)$-dimensional spacetime with the following metric
\begin{align}\label{HDS}
    \dd\tilde s^2=-\dd\tilde t^2+(\tilde A\tilde t)^2 \dd\theta^2+\sum_{1\leq i\leq d-1}(\dd x^i)^2\,.
\end{align}
where $\tilde A=\frac{A^{d-1}(d-2)}{2\pi(d-1)}$ is a constant.
After compactification, we find that the action picks up a volume factor of $2\pi\tilde A\tilde t$. Therefore, to remove this factor, we rescale the metric and define the lower dimensional metric $g_{\mu\nu}=(2\pi\tilde A\tilde t)^{2/(d-2)}\tilde g_{\mu\nu} $. After doing this, the lower dimensional metric takes the form
\begin{align}
    \dd s^2=(2\pi\tilde A\tilde t)^{\frac{2}{d-2}} \left(-\dd\tilde t^2+\sum_{1\leq i\leq d-1}(\dd x^i)^2 \right)\,.    
\end{align}
The proper time $t$ of the comoving observer is $t=\frac{d-2}{d-1}(2\pi\tilde A)^\frac{1}{d-2}\tilde t^\frac{d-1}{d-2}$. We find that the scale factor in the new coordinates is given by 
\begin{align}
    a(t)=A \, t^\frac{1}{d-1}\,.
\end{align}
Therefore, the kination spacetime resulting from decompactification of one dimension uplifts to a higher-dimensional spacetime with metric \eqref{HDS}. Let us define $T=\tilde t\cosh (\tilde A\theta)$ and $X=\tilde t\sinh (\tilde A\theta)$. In the new coordinate system, the metric takes the form 
\begin{align}
    \dd \tilde s^2=-\dd T^2+\dd X^2+\sum_{1\leq i\leq d-1}(\dd x^i)^2\,.
\end{align}
This corresponds to the higher-dimensional Milne space modulo a transformation that shifts $\theta$ by $2\pi$.  The shift  acts on the Minkowski coordinates,  
\begin{align}
    (T,X)\rightarrow \left(\cosh(2\pi\tilde A)T+\sinh(2\pi\tilde A)X,\sinh(2\pi\tilde A)T+\cosh(2\pi\tilde A)X\right)\,.
\end{align}
This is nothing other than a boost by rapidity $\beta=2\pi \tilde A$. For the specific case where the higher-dimensional theory is 11-dimensional supergravity, the spacetime was studied in \cite{Khoury:2001bz} as a potential candidate to realize bounce in string theory.

\subsection{Case III: $\lim_{t\rightarrow\infty}\frac{\dot\phi^2}{H^2}\rightarrow 0$}

Suppose we have another contribution to the energy density that dominates  and drives the cosmology faster than kination in the infinite distance limit such that 
\begin{align}
a(t)\sim t^p\,,
\end{align}
where $p>\frac{1}{d-1}$. If $V\sim |H\dot{\phi}|$, then we have
\begin{align}
    te^{-\lambda\phi}\sim |\dot{{\phi}}|\quad\Rightarrow\quad t^2\sim e^{\lambda\phi}\quad\Rightarrow\quad \dot{\phi}^2\sim t^2\sim H^2\,,
\end{align}
which contradicts our assumption of $|\dot{\phi}|\ll H$. 

Alternatively, if   $V\gg |H\dot{\phi}|$  and  the scalar potential is exponential in the infinite distance limit such that  $V\sim |V'|$,   this implies $|V'|\gg |H\dot{\phi}|$. Therefore, the second term in the  equation of motion for $\phi$ 
\begin{align}\label{eomsf}
\ddot{\phi}+(d-1)H\dot{\phi}+V'=0\,,
\end{align}
is negligible. Now we have  
\begin{align}
\ddot{\phi}\simeq -V'\sim V\gg H\dot{\phi}\quad\Rightarrow\quad \frac{\dd\dot{\phi}}{\dot{\phi}}\gg \frac{\dd t}{t}\,.
\end{align}
Integrating the above inequality leads to 
\begin{align}
    \dot{\phi}\gg t \gg H\,,
\end{align}
which again contradicts our assumption of $|\dot{\phi}|\ll H$. 

Having ruled out $V\sim |H\dot{\phi}|$ and $V\gg |H\dot{\phi}|$, we are left with $|V'|\sim V\ll |H\dot{\phi}|$. In that case, the third term in the equation of motion \eqref{eomsf} is negligible and we find,
\begin{align}
\ddot{\phi}+(d-1)H\dot{\phi}\simeq0\,.
\end{align}
Substituting $H\sim p/t$ in the above equation leads to 
\begin{align}
\ddot{\phi}\sim -\frac{p(d-1)}{t}\dot{\phi}\quad\Rightarrow\quad \phi\sim t^{1-p(d-1)}\,.
\end{align}
Since $p>1/(d-1)$, we can see that the scalar field does not diverge at future infinity but, instead, converges to a finite value. With that, the value of a scalar potential also converges to a constant. A positive scalar potential will eventually be dominant and our assumption that $V\ll H^2$ will no longer be valid. Therefore, we conclude that no cosmology with positive scalar potential falls into this category at future infinity.

\subsection{Summary of Results}\label{SR}

Let us summarize the results of this section. We have observed that spatially flat decelerated expansions in the weak coupling limits of moduli space satisfy  
\begin{align}
    \Lambda_s\lesssim a^{-\alpha},
\end{align}
where \( \alpha \geq 1 \). Apart from the two specific cases discussed below where $\alpha=1$, the universe requires spatial curvature to avoid violating \eqref{ME} and, consequently, to prevent a breakdown of EFT in the future. Notably, the presence of negative spatial curvature always ensures that the inequality \eqref{ME} is satisfied in the asymptotic future.  To understand why, recall that in the presence of negative spatial curvature, the scale factor at future infinity always scales as \( a(t) \sim t \). Thus, inequality \eqref{ME} takes the following form:  
\begin{align}
    \Lambda_s\gtrsim a^{-1}\sim t^{-1}\sim H\,.
\end{align}
However, this inequality is always trivially satisfied since the Hubble parameter must always remain below the quantum gravity cut-off \cite{vandeHeisteeg:2023uxj}.

There are three ways of avoiding the EFT breakdown of the type we described:

\begin{itemize}
    \item The background solution includes a negative spatial curvature which is the only robust resolution that avoids the EFT breakdown in all infinite distance limits.
    \item The scalar field solution is finely tuned such that its trajectory is along a pure decompactification direction (i.e. non-vanishing string coupling) corresponding to a higher-dimensional supersymmetric theory compactified on either a negatively curved internal geometry or a single extra dimension.
    \item The background solution is kept spatially flat but is {\it exactly} FRW with no perturbations. Such solutions are not suited for cosmology given that they are unstable to perturbations. However, this does not invalidate their use in the literature as fine-tuned states to explore the consistency of the Swampland principles.
\end{itemize}

A different and more speculative possibility is that even though the EFT breaks down, the background is still a consistent but non-perturbative quantum gravity solution. For instance, if the physical momentum of a perturbation exceeds the string scale at future infinity while the string coupling vanishes, the background may still admit a consistent description as a weakly coupled worldsheet theory.\footnote{We thank Cumrun Vafa for pointing out this possibility.} In this case, the string theory is weakly coupled while the worldsheet theory is strongly coupled due to the breakdown of $\alpha'$ expansion. For example, one can still define the spacetime metric by relying on the worldsheet CFT as the coefficient of the kinetic term $\partial X^\mu \bar{\partial} X^\nu$. The breakdown of EFT may not create a problem for defining non-perturbative equations of motion which in string theory are given by the vanishing of the worldsheet beta function. However, whether a smooth metric with desirable asymptotic behavior solves such non-perturbative equations is unclear. 

\section{Bounds on Accelerated Expansion in the String Landscape}\label{sec:6}

In this section, we demonstrate that the crucial role played by spatial curvature in ensuring the consistency of EFT at future infinity has important implications for accelerated expansion at finite times. 

Consider the following scenario: Suppose the universe undergoes a period of quasi-de Sitter accelerated expansion, which then transitions into a decelerating expansion driven by a tracking solution in the asymptotic regime of the scalar field space with exponential potentials. If the period of accelerated expansion is excessively long, the spatial curvature contribution becomes highly suppressed, requiring a significant amount of time to re-emerge and play a crucial role. This raises the possibility that the nearly spatially flat expansion of the universe might violate the inequality \eqref{ME}. Consequently, we expect an upper bound on the duration of accelerated expansion. 

\begin{figure}
    \centering
    \includegraphics[width=\linewidth]{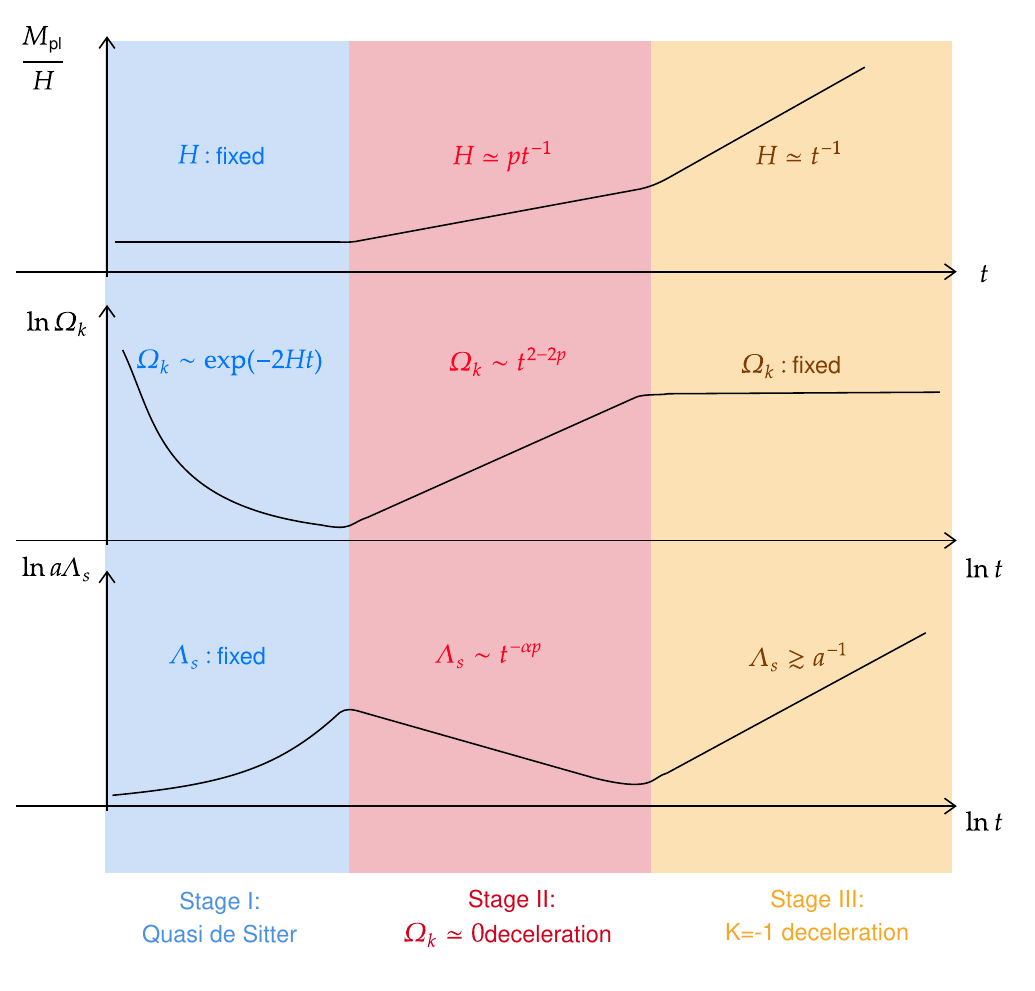}
    \caption{Evolution of the Hubble radius, spatial curvature, and scale factor expressed in units of the quantum gravity cut-off length scale, across the three stages of the cosmological toy model. Stage I is a TCC violating phase of accelerating expansion. Stage II is a decelerating power-law expansion phase given by the spatially flat attractor solution with negligible spatial curvature. Stage III describes a decelerating power-law expansion in which $\Omega_k$ has attained its attractor value.}
    \label{Epochs}
\end{figure}

To simplify the evolution, we divide it into three stages, as shown in Figure \ref{Epochs}.

\begin{itemize}
    \item \textbf{Stage I}: A quasi-de Sitter accelerated expansion with $N$ e-folds, during which the Hubble parameter and the species scale remain nearly constant. We denote them as $H$ and $\Lambda_s$, respectively.
    \item \textbf{Stage II}: A spatially flat FRW solution approximated by the tracking solution, where $a(t)\propto t^p$ and $\Lambda_s\sim a^{-\alpha}$. The constants $p\leq 1$ and $\alpha\geq 1$ are determined by the infinite distance limit parameterized by the scalar field.
    \item \textbf{Stage III}: An FRW universe with negative spatial curvature, where the spatial curvature is locked into its attractor value $\Omega_{k,a}$.
\end{itemize}
During Stage I, $\Omega_k$ decreases as $(aH)^{-2}$. Therefore, at the beginning of Stage II, the value of $\Omega_k$ denoted as $\Omega_{k, II}$ satisfies $\Omega_{k, II} < e^{-2N}$. During the decelerated expansion phase, $\Omega_k$ grows as $(aH)^{-2} \propto a^{2/p-2}$ until it reaches $\Omega_{k,a}$ at the beginning of Stage III. Therefore, we have
\begin{align}\label{eq7}
    \Omega_{k,a}<e^{-2N}\left(\frac{a_{\rm III}}{a_{\rm II}}\right)^{\frac{2}{p-2}},
\end{align}
where $a_{\rm II}$ and $a_{\rm III}$ are respectively the scale factor at the beginning of stages II and III.
During stage II, the species scale evolves as $a^{-\alpha}$, leading to
\begin{align}
    \Lambda_{s, \rm III}\sim \Lambda_s \left(\frac{a_{\rm III}}{a_{\rm II}}\right)^{-\alpha},
\end{align}
where $\Lambda_{s,\rm III}$ is the species scale at the beginning of stage III. To ensure that the species scale remains greater than the proper momentum of a Hubble-sized perturbation at the start of Stage II, we impose
\begin{align}
    H\left(\frac{a_{\rm III}}{a_{\rm II}}\right)^{-1}<\Lambda_s\left(\frac{a_{\rm III}}{a_{\rm II}}\right)^{-\alpha} \quad\Rightarrow\quad \frac{a_{\rm III}}{a_{\rm II}}<\left(\frac{\Lambda_s}{H}\right)^{\frac{1}{\alpha-1}}.
\end{align}
Combining this with \eqref{eq7}, we obtain
\begin{align}\label{TCC2}
    N<\frac{1-p}{p(\alpha-1)}\ln\left(\frac{\Lambda_s}{H}\right).
\end{align}
Here, we have omitted the term $-\ln(\Omega_{k,a})$ on the right-hand side, as it is typically order one and much smaller than $\ln(\Lambda_s/H)$. 

This inequality closely resembles the TCC bound on the number of e-folds of accelerated expansion, except for the coefficient. Since $\alpha$ can be precisely equal to one in finely tuned cases, the bound above can be substantially weaker than the TCC, which replaces the coefficient $\frac{1-p}{p(\alpha-1)}$ with 1. However, it robustly establishes a bottom-up argument for an upper bound on the duration of accelerated expansion, showing that it is proportional to $\frac{1}{H}\ln(\frac{M_{\rm pl}}{H})$ rather than a much longer timescale, such as the Poincar\'e timescale $\sim e^{(M_{\rm pl}/H)^{d-2}}$.

Furthermore, we highlight that our new bound \begin{align}
    \frac{a_f}{a_i}>\frac{H_i}{\Lambda_{s,f}}\,,
\end{align}
and the TCC \begin{align}
    \frac{a_f}{a_i}<\frac{\Lambda_{s,i}}{H_f}\,,
\end{align}
together imply the constraint
\begin{align}
    H_i H_f<\Lambda_{s,i}\Lambda_{s,f}\,,
\end{align}
which is trivially satisfied due to the species scale always being smaller than the Hubble scale, as pointed out in \cite{vandeHeisteeg:2023uxj}. Therefore, our inequality remains compatible with the TCC and indeed leads to a version of it in \eqref{TCC2} for periods of accelerated expansion that continuously evolve into decelerated expansion in the future.

We note an important qualification about the implications of our results. Our bounds apply to the duration of the last period of accelerated expansion before the scalar field classically evolves to the asymptotics of the moduli space. Therefore, our bounds can potentially apply to the present dark energy dominated phase. However, if the universe underwent an inflationary phase in our past, it is not subject to our constraints because inflation must have been followed by a radiation dominated phase. Similarly, the bounds would not apply to the dark energy dominated phase if it transitions to a radiation or matter dominated phase. Since our bounds must hold for any trajectory in the string landscape, they could still constrain the inflationary potential given the existence of a classical scalar field cosmology that connects an accelerated expansion sourced by the inflationary potential to the asymptotics of the moduli space. In this way, we place a set of highly non-linear constraints on the interior of the moduli space which echo the findings of \cite{Bedroya:2024zta} that achieved similar constraints by imposing consistency conditions motivated by holography. On the one hand, our constraints may seem weak because they do not apply to most trajectories in a scalar field potential energy landscape, including perhaps the cosmological history of our observable universe if all phases of accelerated expansion are followed by a sustained period of radiation or matter domination. On the other hand, the constraints can be viewed as very strong because in any landscape consistent with quantum gravity, the constraints must apply to all trajectories without exception, which essentially imposes an infinite number of conditions \cite{Bedroya:2024zta}.

\section{Constraints on Contracting Universes}\label{sec:7}

We can generalize the inequality \eqref{ME} to contracting universes. Similar to the case of expanding universes, where, as explained in Section \ref{sec:5}, a superficial trans-Planckian issue in the past direction does not necessarily imply a breakdown of effective field theory (EFT), a superficial trans-Planckian issue in the future direction of a contracting universe likewise does not indicate a breakdown of EFT. However, a breakdown of EFT due to a violation of the condition \( |H_f| \frac{a_f}{a_i} \ll \Lambda_{s,i} \) in the past is prohibited.

We can reformulate our inequality \eqref{ME} in a way that applies equally to both contracting and expanding solutions. Consider an FRW solution evolving between two times, where the greater of the two scale factors is denoted as \( a^+ \) and the smaller as \( a^- \). Furthermore, let the Hubble scale at the time associated with \( a^- \) be given by \( H^- \), and the species scale at the time associated with \( a^+ \) be \( \Lambda^+ \). Then, the following inequality holds:
\begin{align}\label{TSI}
    a^-H^- < \Lambda_s^+ a^+ \,.
\end{align}

Since the inequality \eqref{TSI} also applies to contracting universes, we can extend the arguments of section \ref{sec:6} to contracting universes as well. The outline of the argument is as follows: Suppose we have a contracting universe that began in the infinite past within the weakly coupled regime of the string landscape. As the scalar field evolved toward the interior of the field space, the contraction transitioned from a decelerating phase (slow contraction) to an accelerating phase (fast contraction). The phase of accelerated contraction could not have lasted for too long. If it had, the spatial curvature at the onset of this stage would have been too small, leading to an extended period of nearly flat, decelerated contraction in the weakly coupled region of the string landscape, which would violate the inequality \eqref{TSI}. Thus, we conclude that the inequality \eqref{TCC2} holds for accelerated contractions as well and the number of e-folds during a quasi-de Sitter accelerated contraction phase must satisfy $N<\frac{1-p}{p(\alpha-1)}\ln(\frac{\Lambda_s}{|H|})$. Similar to the expanding case, the constants $p\leq 1$ and $\alpha\geq 1$ are determined by the infinite distance limit in the scalar field space, from which the contracting spacetime originated in the infinite past.

Let us emphasize that the above argument relies on the existence of an infinite direction in time without a breakdown of EFT. For instance, a universe in which the domain of EFT validity is bounded both in the future and the past—whether due to a cyclic nature or the presence of genuine big bang/crunch singularities—does not satisfy our inequalities. We make this assumption motivated by holography, which suggests that observables in a nontrivial quantum gravity theory reside on the asymptotic boundary of spacetime. A recent exploration of the implications of this assumption for cosmological models can be found in \cite{Bedroya:2024zta}.

\section{When is $\left|\frac{\nabla V}{V}\right| \geq \frac{2}{\sqrt{d-2}}$ saturated?}\label{sec:8}

In this section, we take a detour to explore the regions of the string landscape where the inequality
\begin{equation}\label{TCC1}
    \left|\frac{\nabla V}{V}\right| \geq \frac{2}{\sqrt{d-2}}\,,
\end{equation}
can be saturated in the infinite-distance limit of moduli space. This discussion also highlights the power of Swampland principles as organizing tools in the study of string constructions. We argue that unitary string theories which have vanishing tree-level contribution to the scalar potential cannot saturate this inequality. The only claimed example that we are aware of to saturate the bound \eqref{TCC1} is the supercritical string theory mentioned in \cite{Rudelius:2021azq}. However, this is not a valid example given that the usual no-ghost theorem proving the unitarity of critical string theory does not apply to the supercritical case. Here we provide an argument explaining why any corner of the quantum gravity landscape with $\left|\frac{\nabla V}{V}\right| =\frac{2}{\sqrt{d-2}}$
would constitute a counterexample to the Emergent String Conjecture.

From Eq.~\eqref{eq1}, we know that weak coupling limits satisfy
\begin{align}\label{eq5}
   \left|\frac{\nabla V}{V}\right| \times \left|\frac{\nabla \Lambda_s}{\Lambda_s}\right| \geq 2 \frac{\nabla \sqrt{V}}{\sqrt{V}} \cdot \frac{\nabla \Lambda_s}{\Lambda_s} \geq 2 \frac{\nabla m}{m} \cdot \frac{\nabla \Lambda_s}{\Lambda_s}=\frac{2}{d-2}\,.
\end{align}
On the other hand, utilizing the emergent string conjecture, the authors in \cite{vandeHeisteeg:2023uxj} demonstrated that
\begin{align}\label{eq6}
    \left|\frac{\nabla \Lambda_s }{\Lambda_s}\right| \leq \frac{1}{\sqrt{d-2}}\,,
\end{align}
and that saturation occurs in string limits where $\Lambda_s$ represents the mass scale of a fundamental weakly coupled string. 

If $\left|\frac{\nabla V}{V}\right| = \frac{2}{\sqrt{d-2}}$, then both inequalities \eqref{eq5} and \eqref{eq6} must be saturated. From the saturation of Eq.~\eqref{eq6}, we deduce that the direction of steepest descent for $\Lambda_s$ corresponds to a string limit. Similarly, from the saturation of Eq.~\eqref{eq5}, we infer that the gradient of steepest descent for $V$ aligns with that of $\Lambda_s$. Moreover, the saturation of Eq.~\eqref{eq5} implies that the scalar potential is proportional to $m^2$. In a string limit, this corresponds to
\begin{align}
    V \sim M_s^2 M_{\rm pl}^{d-2}\,,
\end{align}
which represents the contribution of the sphere worldsheet to the scalar potential. This contribution vanishes in critical string theories. 

Next, we show that decompactification to dimension $D$ in the large $D$ limit can approach saturation arbitrarily closely. However, within the string landscape, $D$ is bounded above by $11$. Suppose the scalar potential is sourced by the curvature of a negatively curved internal manifold. In this case, the scalar potential is proportional to $m_{\rm KK}^2$, and if the theory decompactifies to $D$ spacetime dimensions, we obtain
\begin{align}
    \left|\frac{\nabla V}{V}\right| = \frac{2\sqrt{D-2}}{\sqrt{(D-d)(d-2)}} > \frac{2}{\sqrt{d-2}}\,.
\end{align}
The larger the value of $D$, the closer the inequality $\left|\frac{\nabla V}{V}\right| \geq \frac{2}{\sqrt{d-2}}$ comes to being saturated. Many of the inequalities such as $|\frac{\nabla m}{m}|\geq \frac{1}{\sqrt{d-2}}$ and $|\frac{\nabla\Lambda_s}{\Lambda_s}|\leq \frac{1}{\sqrt{d-2}}$ are also saturated in the large-$D$ limit. As also proposed in \cite{Bonnefoy:2020uef}, there appears to be a more fundamental meaning to the large-$D$ limit as it mimics a string limit for many Swampland inequalities.

\section{Conclusions}

We have identified a novel constraint on the validity of EFT in expanding spacetimes: the species scale must not decrease so rapidly that Hubble-sized perturbations exceed it. This imposes the condition $H_i \frac{a_i}{a_f} \ll \Lambda_{s,f}$ on any expanding FRW universe. While similar in form to the TCC, this condition is fundamentally different, as a violation of the TCC does not necessarily lead to an EFT breakdown, whereas a violation of our constraint does. 

The key distinction lies in the fact that Hubble friction turns into anti-friction as we extrapolate backward in time. This anti-friction resolves trans-Planckian issues which appear in the past but not the EFT breakdowns that appear in the future. In particular, Hubble friction in expanding spacetime suppresses the amplitudes of superhorizon modes in the future and amplifies them in the past. The amplifications of superhorizon modes in the past can increase the average wavelength of a perturbation consisting of a linear combination of modes and smooth it out, while the suppression of such modes in the future makes it impossible to resolve the EFT breakdown of the kind that we have discussed here.

One of the most striking implications of our work is that spatially flat FRW solutions in the string landscape generally violate the condition required for EFT validity. \textit{ An important implication is that most spatially flat FRW solutions that are studied in the literature as cosmological backgrounds are invalid for that purpose.} The only exceptions where a flat FRW solution in the string landscape can remain stable occur in higher-dimensional supersymmetric theories compactified on $i$) a negatively curved internal geometry or $ii$) an extra spatial dimension. 

\textit{Another striking implication is that a sufficiently large negative spatial curvature always avoids the EFT breakdown.} In these cases, we derived an upper bound on the duration of quasi-de Sitter expansion before rolling toward the asymptotic field space:
\begin{align}
    \tau<\frac{1-p}{p(\alpha-1)}\frac{1}{H}\ln\left(\frac{\Lambda_s}{H}\right),
\end{align}
where $p \leq 1$ and $\alpha \geq 1$ characterize the scalar field trajectory in the infinite distance limit defined in Sec.~\ref{sec:6}. Remarkably, this result reproduces the TCC upper bound up to a numerical factor, providing a bottom-up argument that extends to the interior of the string landscape.

Finally, we note that these conclusions were derived from $\frac{\nabla m}{m}\cdot\frac{\nabla \Lambda_s}{\Lambda_s}=\frac{1}{d-2}$ and therefore offer some insight about the physical meaning of this identity which has so far lacked any motivation outside of string theory.

\section*{Acknowledgements}

We are grateful to Georges Obied, Savdeep Sethi, and Cumrun Vafa for valuable discussions, and we thank David Andriot for helpful comments on the draft. AB is supported in part by the Simons Foundation grant number 654561 and by the Princeton Gravity Initiative at Princeton University. PJS is supported in part by the Department of Energy grant number DEFG02-91ER40671 and by the Simons Foundation grant number 654561. 

\bibliographystyle{unsrt}
\bibliography{References}

\end{document}